# Regional Compartmentalization in Multienzyme-related Biomaterials System

Zhexu Xi[(1)]

Bristol Center of Functional Nanomaterials, School of Physics, University of Bristol[(1)]

**Abstract:** Multienzyme cascaded reactions are widely utilized because they can generate value-added biomaterials and biodevices from simple raw materials. However, how to promote the catalytic efficiency and synergistic effect of the multienzyme system is proved to be a challengeable point. Recent discovery repeatedly emphasized the strategy of assembled multienzyme complexes or forming subcellular compartments for spacial optimization. This highly ordered and tunable organization contributes to various biochemical processes. This dissertation focuses mainly on analysis and progresses in this cascaded strategy, regarding the feasibility of regional compartments for natural or artificial biochemical reactions in vivo and vitro, simultaneously.

## 1. Introduction

Recently, enzymes are increasingly used in cascaded reactions because of high selectivity to promote a biofriendly synthetic environment, like multienzyme cascaded reactions. Compared with traditional multistep sequential biosynthesis methods, enzyme cascades can relieve the bulk diffusion from unstable or toxic intermediates and avoid the intermediary product purification[1,2].

Additionally, sequential cascaded reactions in vivo are more attractive in the stage of the multienzyme assembled system. They can effectively eliminate product inhibition, improve cofactor regeneration and work without purification and immobilized enzymatic catalysts. Common methods of reaction pathways contain two following steps[1,3]. One is based on optimization and modification of natural intermediate metabolism by gene knockout or overexpression in the early period. Lonza et al.[4] found a biological process in natural metabolic pathways to synthesize L-carnitine from 4-butylamine. They got a strain of microorganisms that can degrade 4-butylamine by L-carnitine co-enzyme A and Krebs cycle to knock out L-carnitine deaminase, functioning as the first step to catalyze the degradation. Next is to introduce heterologous enzymes to optimize cascaded metabolism and synthesize non-natural products. Ro et al.[5] founded a novel high-production process of cyanuric acid by sorting out the engineered

Mevalonate pathway from Saccharomyces cerevisiae. The last mainly focuses on the design of a non-natural metabolic route to integrate enzymes from different origins into the same cellular space for expression. This approach can meet the demands in a large production of the enzymes with an extensive substrate spectrum, which may achieve co-expression in the same space like multienzyme coupling reactions[6].

Accordingly, many techniques in vitro spark wide attention due to the vitro multiezyme reactions' improvements with a scaling effect in the sequential signal transfer. The vitro multiezyme process has a simple and easily controllable structure compared to the vivo traditional methods. Admittedly, with a mature strategy reconstructing of a multienzyme system or pseudo-natural synthesis by artificial design, it can freely choose initial substrates without effect from metabolic intermediates. Ox-red enzyme-related cascaded system is one of the typical examples[7].

During the developing but tortuous period of research, the more cascade numbers it reached, the more uncertain parameters needed to be taken into account, like enzymatic stability, intermediate transfer and substrate inhibition in vitro, or rate-limiting enzymes, competitive inhibition and diffusion of toxic intermediates in vivo[8,9]. Technology in artificial multienzyme simulation confronts a challengeable problem: establishing an optimized spacal organization for enzyme assemblies to handle the increasingly sophisticated requirements in metabolic engineering.

Natural pathway for cascaded bioreactions is to make two or more active sites of enzymes close to each other to accelerate the enzymatic reactions, which indicates a regional optimization for spacal arrangement and energy barrier. Based on this, we may talk about a perfect structural assembly with the dynamic interaction in the multienzyme cascaded reactions as well as its signs of progress and applications.

## 2. Brief review on substrate channeling with multienzyme assemblies

As mentioned above, a common way of compartmentalization is called substrate channeling interaction, which means a metabolic transferring of intermediates from one enzyme directly to another without its bulk diffusion into the solution. Based on its multi-function, it has been widespread in the process of enzymatic reactors and bioengineering. In nature, there are three kinds of pathways occurring the construction of spacal organization: One, which occurs in the pyruvate dehydrogenase complex, is by a substrate attached to a flexible arm that moves between several active sites in a protein complex; Another possibility is by two active sites connected by a tunnel where the enzymes and the substrate moving in the protein, such as tryptophan synthase; A third way is by a charged region on the surface of the enzyme acting as a "electrostatic highway" to guide a substrate that has the opposite charge from one active site to another, like bifunctional enzyme dihydrofolate reductase-thymidylate synthase.

Like the Calvin cycle, biosynthesis of amino acids and microbodies, many primary metabolic pathways can be more effective with the secondary metabolic factors, majorities known as multienzyme assembled complexes. Tryptophan synthase, existing in the form of (αβ)$_2$ complexes as a domain of homologous proteins, catalyzes the last two tryptophan synthetic methods. Anderson et al.[10] discovered the evidence of the indole channel through steady-state experiments. Only a very small amount of indole was detected during the synthesis of L-tryptophan, which was about 1% of the concentration of D-3-glyceraldehyde, along with undetectable synthetic delay time of L-tryptophan. The results of kinetic experiments show that indole rapidly moves from the active site of the α subunit to β and rapidly converts to tryptophan on the β subunit. This research demonstrates an ordered hydrophobic tunnel between 2 subunits, whose allosteric effect may further regulate the active space[11]. According to the discovery, roles in substrate channeling become clear: a) becoming increasingly sensitive to environmental parameters; b) protecting the unstable intermediates and cofactors; c) relieving the toxicity diffusion of metabolism; d) improving cellular dissolving capacity due to insoluble intermediates; e) promoting catalytic efficiency.

Thus, spatial organization plays a significant role in assembled strategy design, but more advanced and pointed optimization to guarantee the highly ordered catalytic efficiency needs further exploration.

### 3. Strategies of the spatial organization by assembling enzymes

More optimized artificial assemblies need introducing to build an efficient substrate highway, which ensures the outstanding catalytic conversion. Presently, there are several successful multienzyme assembled methods, including co-immobilization, protein fusion assembly, protein cross-linked collocation assembly, and multi enzymatic assembly of macromolecular scaffolds.

### 3.1 Co-immobilization

Inspired by the natural multienzyme complexes, more attention is fixed on the multienzyme co-immobilized system. It can improve the enzymatic stability, make bienzymes closer to each other and increase the concentration of substrates around the enzyme to promote the enzymatic efficiency[12]. The multi-enzyme system is also suitable for the in-situ coenzyme regeneration reaction process, greatly reducing the cost of large-scale biocatalytic reactions.

Based on the methods of single enzyme techniques, the common form of a co-immobilized system contains immobilization on carriers and cross-linked aggregation of multienzymes, while the modern multienzyme ways are by multi-site covalent bonding, physical embedding, physical adsorption, site-specific affinity interaction, DNA directed self-assembly immobilization, and cross-linked enzyme aggregation with their combinations.

However, there are still many challenges in the field of multienzyme co-immobilization technology. First, the structure and function of a single enzyme should be optimized to maintain the activity and stability of each enzyme. Secondly, the carrier should have great compatibility to not affect the enzyme activity, also to guarantee the maximal loading of enzymes. Finally, the position of different enzymes in the carrier is effectively controlled, thereby reducing the limit of the inherent diffusion[13].

**3.2 Protein fusion assembly**

Multienzyme fusion refers to integrating of multienzymes by gene fusion to generate a short-chain chimeric protein linked with a short peptide. As many results reveal, multifunctional enzymes show greater activity than single enzymes, and catalytic efficiency grows with the increasing concentration of substrates[14]. Multi-fusion technology can keep the independence of every single enzyme in function and steric structure, and make them closely associated with each other, defined as proximity effect, based on the result of small-angle scattering analysis[15].

With the booming of gene engineering, cascaded multienzymes are more widely utilized and easily spatial-modified using multi-terminal fusion and other useful bio-related strategies. In view of the goal of constructing a fusion protein with expected characteristics, there is a high uncertainty because present synthesis counts primarily on subjective experience. It is vital and urgent to understand micro-level information of fusion proteins, like spacial orientation, interactions and dynamics in folding and allosteric regulation. A systematic informatics tool need simultaneously keep pace for exact simulation and rational optimal design.

**3.3 Protein cross-linked collocation assembly**

The hybrid cross-linked enzyme is one of the most representative examples of chemical cross-linking methods for multi-enzyme assembly. The methods use single-enzyme cross-linked enzyme aggregates to effect the cross-linking aggregation of two or more enzymes. For example, cross-linked enzyme aggregates, which are formed by co-crosslinking S-selective oxynitrilase and nitrilase catalyze, converse benzaldehyde S-mandelic acid. It can extraordinarily reduce the $K_m$ value during the catalytic reaction and uniquely eliminate the enzyme-inhibition effects of nonspecific reactions.

Such aggregations are synthesized by disordered cross-linking of surface groups, which illustrates an uncertainty of spatial approach to each other, so novel chemo-selective methods have been identified for protein aggregations to realize the heterologous assembly such as sulfhydryl-sulfate exchanging reactions to form a new peptide bond[16].

**3.4 Roles of protein scaffolds**

In cellulosome, the existence of protein scaffolds, containing cellulose-binding proteins without catalytic reactivity and many fibronectins, makes contributions to combination with a domain of anchoring the protein to realize the assembly of hydrolase. The most representative scaffold is $(GBD)_x$-$(SH3)_y$-$(PDZ)_z$ by Dueber[17], which functioned well in the assembly of cascaded enzymes during metabolic pathways. The optimal output is achieved by regulating the ratio of x, y and z values.

On the basis of this, interactions among multi-proteins gradually play an important part in constructing scaffolds. For instance, SP1 protein originates from homomultimeric proteins in poplars, with a great high-temperature, pH and solvent resistance. Heyman et al. fused glucose oxidase with SP1 and expressed it in E. coli. Finally, the enzyme molecule was assembled into enzyme-labelled nanotubes by SP1 scaffolds. Each nanotube contained hundreds of enzyme molecules with an outstanding catalytic activity. Then, they accomplished the fusion between Coh and the SP1 gene and achieved the soluble expression of the fusion protein Coh-SP1 in E. coli so that there are 12 sites for co-binding proteins(Doc) connected to the Coh-SPl per molecule. When the cellulose derived from Thermobifida fusca was fused with Doc and then mixed with Coh-SP1 in vitro, the cellulose-based scaffold was obtained, whose catalytic activity was remarkably improved[18].

**3.5 Roles of DNA scaffolds**

DNA stands for the expressing carrier of various molecular messages, storing and transmitting genetic instructions. With a unique hereditary property, DNA polymers can be extensively used in rational design by changing the base sequences.

Fu et al.[19] constructed a double-crosslinked scaffold-mediated complex containing swinging arm and multienzymes. The 6-phosphoglucose deaminase (G6pDH) and malate deaminase (MDH) were modified with a pre-encoded oligonucleotide that specifically binds the DNA scaffold so that G6pDH and MDH can assemble spontaneously at a certain distance to the DNA. In order to enhance the substrate channeling between G6pDH and MDH, they used NAD+ modified by the oligonucleotide to bind a site on the scaffold between 2 enzymes to form a multienzyme complex that co-enzymes can shuttle back and forth. Regarding the swinging arm, the double-enzyme reaction enhances its specificity among numbers of unmeasurable reaction systems, besides its boosted reactivity, as the results depicted.

Moreover, the combination between DNA and its conjugated protein can also involve in the DNA-scaffold-mediated methods to assemble into multienzyme complexes. Usually, different zinc-finger-combined DNAs are inserted into the plasmids with a fusion between domains of zinc finger proteins and scaffold plasmids simultaneously[20]. After their co-expression in the E. coli, cascaded enzymes can be anchored in the corresponding position under the driving interactions between domains of zinc finger proteins and their binding sites. By optimal changing the number and alignment of the zinc finger binding sites, greater multienzymatic efficiency may be achieved.

**4. Comments on compartmentalization in metabolic pathways**

It is natural to accomplish metabolism by regionally separated enzymes from vesicles, membranes, cellular organelles, or non-membrane bound cellular granules in the cytoplasm. Substrate channeling effect has long been identified to promote metabolic efficiency by multienzyme complexes, but over four sequential cascaded reactions in the tunnel may be structurally limit and more challengably operated. In recent years, highly controlled spatial aggregations are proved to be well tuned with the media of cluster for substrate channeling. Under this circumstance, effective and ordered collocation of sequential enzymes may primarily contribute to the catalytic reactivity instead of associated active bindings.

There is a lack of a direct signal from the formation of assembled complexes to the metabolic flux enhancement, so the cascaded reactions in metabolism may contain more sophisticated mechanisms for multifunctional spacial assemblies[21]. In other words, the spatial organization is only suitable for a specific bio-compatible environment with single enzymes of given pathways, which shows little potential functional diversities.

**5. Outlook**[22]

a. Multi-protein fusion lacks theoretical guidance because of the lack of understanding of interactions between intracellular structures with their relationship with functions. Like improving the soluble expression level of the fusion protein and facilitating the impact on the folding protein, more deeper protein-related information needs to be explored.

b. How to accomplish the rational design of enzymatic assemblies is still a big challenge, indicating that more bioinformatic methods or tools need to keep pace with cascaded reactions. By establishing a homologous structural model of intracellular multienzymes, a suitable binding site can be easily found to achieve the exact control of the spatial orientation of enzymes on the surface for efficient immobilization.

c. For substrate channeling mechanism, contact via large numbers of sequential enzymatic reactions is hard to accomplish, which break the common sense that substrate channeling can promote the catalytic efficiency and relieve the bulk diffusion from toxic or unstable intermediates. Accordingly, the entire mechanism for assembled complexes needs further study.

d. For the precise control, more bioreaction mechanisms need exploring, like energy barrier change during catalysis and allosteric effect. To reveal the mechanisms of controlled assembly of protein complexes, more advanced strategies of dynamic simulation of conformational changes in the assembling process concerning multi-level interactions should be developed.

e. Supramolecular interactions between active sites on different enzymes can be taken into key consideration, for the system is easily tunable under optimal regulation and form micro-partitions when multi-aggregation. Additionally, it can be a significant method for assemblies to eliminate reducing enzymatic activity without purification, which can generate the ordered morphology.

**References**

[1] F. Lopez-Gallego, C. Schmidt-Dannert. Multi-enzymatic Synthesis [J]. *Curr. Opin. Chem. Biol.,* 2010, 14(2): 174-183.

[2] M. Mathesh, J. Q. Liu, C. J. Barrow, W. R. Yang. Graphene-Oxide-Based Enzyme Nanoarchitectonics for Substrate Channeling [J]. *Chem. Eur. J.,* 2016, 22: 1-9.

[3] X. Gao, S. Yang, C. C. Zhao. Artificial Multienzyme Supramolecular Device: Highly Ordered Self-assembly of Oligomeric Enzymes in Vitro and in Vivo [J]. Angew. Chem. Int. Ed., 2014, 53: 14027-14030.

[4] M. Wieser, K. Heinzmann, A. Kiener. Bioconversion of 2-cyanopyrazine to 5-hydroxypyrazine-2-carboxylic Acid with Agrobacterium sp. DSM 6336 [J]. *Appl. Microbio. Biot.,* 1997, 48(2): 174-176.

[5] D. -K. Ro, E.M. Paradise, M. Ouellet, et al. Production of the Antimalarial Drug Precursor Artimisinic Acid in Engineered Yeast [J]. *Nature.,* 2006, 440(7086): 940-943.

[6] M. K. Akhtar, N. J. Turner, P. R. Jones, et al. Carboxylic Acid Reductase is a Versatile Enzyme for the Conversion of Fatty Acids into Fuels and Chemical Commodities [J]. *P. Natl. Acad. Sci. USA.,* 2013, 110(1): 87-92.

[7] W. Liu, P. Wang. Cofactor Regeneration for Substainable Enzymatic Biosynthesis [J]. *Biotechnol. Appl.,* 2007, 25(4): 369-384.

[8] R.J. Ellis. Macromolecular Crowding: Obvious but Underappreciated [J]. Trends Biochem. Sci., 2001, 26: 597−604.

[9] I. Yu, T. Mori, T. Ando, et al. Biomolecular Interactions Modulate Macromolecular Structure and Dynamics in Atomistic Model of a Bacterial Cytoplasm. *eLife 5*, 2016: e19274.

[10] S. Ahmed, S. Ruvinov, A. Kayastha, E. Miles. Mechanism of Mutual Activation of the Trytophan Synthase of Alpha and Beta Submits. Analysis of the Reaction Specificity and Substrate-induced Inactivation of Acitve Site and Tunnel Mutants of the Beta Subunit [J]. *J. Biol. Chem.,* 1991, 266(32): 21548-21557.

[11] M. F. Dunn. Allosteric Regulation of Substrate Channeling and Catalysis in the Trophan Synthase Bienzyme Complex [J]. Arch. Biochem. Biophys., 2012, 519(2): 154-166.

[12] R.C. Rodrigues, C. Ortiz, A. Berenguer-Murcia, et al. Modifying Enzyme Activity and Selectivity by Immobilization [J]．*Chem. Soc. Rev.*, 2013, 42(15): 6290-6307．

[13] F. Jia, B. Narasimhan, S. Mallapragada. Materials-based Strategies for Multi-enzyme Immobilization and Co-localization: a Review [J]. *Biotechnol. Bioeng.*, 2014, 111(2): 209-222.

[14] P. Ljungcrantz, H. Carlsson, M. O. Mansson, et al. Construction of an Artificial Bifunctional Enzyme, β-galactosidase/galactose Dehydrogenase, Exhibiting Efficient Galactose Channeling [J]. *ACS. Biochem.* 1989, 28(22): 8786-8792.

[15] L.F. Ribeiro, G.P. Furtado, M.R. Lourenzoni, et al. Engineering Bifunctional Laccase-xylanase Chimeras for Improved Catalytic Performance [J]. *J. Biol. Chem.*, 2011, 286(50): 43026-43038.

[16] C.P. Hackenberger, D. Schwarzer. Chemoselective Ligation and Modification Strategies for Peptides and Proteins [J]. Angew. Chem. Int. Edit. 2008, 47(52): 10030-10074.

[17] J. E. Dueber, G. C. Wu, G. R. Malmirchegini, et al. Synthetic Protein Scaffolds Provide Modular Control over Metabolic Flux [J]. Nat. Biotechnol. 2009, 27(8): 753-759.

[18] R. Thiruvengadathan, V. Korampally, A. Ghosh, et al. Nanomaterial Processing Using Self-assembly-bottom-up Chemical and Biological Approaches [M]. Rep. Prog. Phys. 2013, 76(6): 066501.

[19] J. Fu, Y.R. Yang, A. Johnsonbuck, et al. Multi-enzyme Complexes on DNA Scaffolds Capable of Substrate Channelling with an Artificial Swinging Arm [J]. Nature Nanotech. 2014, 9(7): 531-536.

[20] R. J. Conrado, G. C. Wu, J. T. Boock, et al. DNA-guided Assembly of Biosynthetic Pathways Promotes Improved Catalytic Efficiency [J]. *Nucleic Acids Res.* 2012, 40(4): 1879.

[21] I. Petrovska, E. Nuske, M.C.Munder, et al. Filament Formation by Metabolic Enzymes is a Specific Adaptation to an Advanced State of Cellular Starvation. *eLife 3,* 2014, e02409.

[22] S. Z. Wang, Y. H. Zhang, H. Ren, et al. Strategies and Perspectives of Assembling Multi-enzyme Systems [J]. *Crit. Rev. Biotechnol n.* 2017, 37(8): 1024-1037.